 \documentclass[smallabstract,smallcaptions]{dccpaper}

\usepackage{amsmath,stmaryrd}
\usepackage{color}
\usepackage{dsfont}
\usepackage{url}
\usepackage{graphicx}
\usepackage{subfigure}
\usepackage{soul}
\usepackage{amsthm}

\usepackage{caption}
\usepackage{lipsum}
\usepackage{algorithmic}

 \usepackage{amssymb}
\usepackage{soul}
\usepackage{mathrsfs}

\usepackage{setspace}

\usepackage{mathtools}

\newfont{\bbb}{msbm10 scaled 500}

\newfont{\bb}{msbm10 scaled 1100}

\def\<{\langle}
\def\>{\rangle}

\newlength{\figurewidth}
\newlength{\smallfigurewidth}

\newtheorem{prop}{Proposition}

\begin{document}

\title
{\large
\textbf{K-means Algorithm over Compressed Binary Data}
}

\author{Elsa \textsc{Dupraz} \\
\small IMT Atlantique, Lab-STICC, UBL, Brest, France
}

%

\maketitle
\thispagestyle{empty}

\begin{abstract}
 We consider a network of binary-valued sensors with a fusion center. The fusion center has to perform K-means clustering on the binary data transmitted by the sensors.
In order to reduce the amount of data transmitted within the network, the sensors compress their data with a source coding scheme based on binary sparse matrices.
 We propose to apply the K-means algorithm directly over the compressed data without reconstructing the original sensors measurements, in order to avoid potentially complex decoding operations.
 We provide approximated expressions of the error probabilities of the K-means steps in the compressed domain.
 From these expressions, we show that applying the K-means algorithm in the compressed domain enables to recover the clusters of the original domain.
 Monte Carlo simulations illustrate the accuracy of the obtained approximated error probabilities, and show that the coding rate needed to perform K-means clustering in the compressed domain is lower than the rate needed to reconstruct all the measurements.
 \end{abstract}

\section{Introduction}
Networks of sensors have long been employed in various domains such as environmental monitoring, electrical energy management, and medicine~\cite{yick08CN}. 
In these networks, inexpensive binary-valued sensors are successfully used in a wide range of applications, for example in traffic control in telecommunication systems~\cite{wang03AC}, self-testing in nanonelectric devices~\cite{colinet10AC}, or activity recognition on home environments~\cite{ordonez13S}. 
In this paper, we consider a network of $J$ sensors that transmit their data to a fusion center.
The fusion center may realize complex data analysis tasks by aggregating the sensors measurements and by exploiting the diversity of the collected data. 
This paper considers clustering as a particular data analysis task that consists of separating the data into a given number of classes with similar characteristics.
One of the most popular clustering methods is the K-means algorithm~\cite{jain10PRL} due to its simplicity and its efficiency.
The K-means algorithm groups the $J$ measurement vectors into $K$ clusters so as to minimize the average distance between vectors in a cluster and the cluster center.
K-means algorithms were proposed for real-valued measurements~\cite{jain10PRL} but also for binary measurements~\cite{huang98DMKD}. 

In our context, the $J$ sensors should send their measurements to the fusion center in a compressed form in order to greatly reduce the amount of data transmitted within the network.
The standard distributed compression framework~\cite{xiong04SP} considers that the fusion center has to reconstruct all the measurements from all the sensors.
However, in the aforementioned applications~\cite{wang03AC,colinet10AC,ordonez13S}, the objective of the fusion center is not to reconstruct the measurements, but to perform a given learning task over the data. 
This is why in order to avoid useless and potentially complex decoding operations, we  propose to perform K-means directly over the compressed data. This approach raises three questions: 
(i) How should the data be compressed so that the fusion center can perform K-means without having to reconstruct all the measurements? (ii) How good is clustering over compressed data compared to clustering over the original data? (iii) Is the rate needed to perform K-means lower than the rate needed to reconstruct all the data? 

Regarding the first two questions,~\cite{boutsidis15IT,keriven16arxiv} consider real-valued measurement vectors compressed from Compressed Sensing (CS) techniques, and show that it is possible to apply the K-means algorithm directly in the compressed domain.
Other than K-means, detection and parameter estimation can be applied over compressed real-valued data~\cite{davenport10SP,zebadua15arxiv}, but also over compressed binary data~\cite{Wang12CSVT,dupraz14Com}.
However, none of these works consider the K-means algorithm over compressed binary data.
Regarding the third question, to the best of our knowledge, the K-means algorithm has not been studied yet in the information theory framework.

Binary data may come either from binary-valued sensors, or from the binary representation of quantized real measurements. 
Here, we propose a clustering algorithm that applies directly over compressed binary data. 
Our approach is based on sparse binary matrices for compression and on K-means in order to cluster the compressed binary vectors.
In order to validate this approach, we further propose a theoretical analysis of the performance of the K-means algorithm in the compressed domain. 
We in particular derive analytical approximated expressions of the error probabilities of each of the two steps of the K-means algorithm.
The theoretical analysis shows that the K-means algorithm in the compressed domain permits to successfully recover the clusters of the original domain.
Monte Carlo simulations confirm the accuracy of the obtained approximated error probabilities. 
We also show from Monte Carlo simulations that the practical rate needed to perform K-means over compressed data is lower than the rate needed to reconstruct all the measurements.


The outline of the paper is as follows.
Section~\ref{sec:system} presents the statistical model we consider for the measurement vectors and describes the source coding technique that will be used in our system.
Section~\ref{sec:method} introduces the K-means algorithm in the compressed domain.
Section~\ref{sec:perf} proposes the theoretical analysis of the two steps of the K-means algorithm. 
Section~\ref{sec:results} presents our Monte Carlo simulation results.


\section{System Description}\label{sec:system}
In this section, we first introduce our notations and assumptions for the binary measurement vectors collected by the sensors.
We then present the source coding technique that is used in the system.

\subsection{Source Model}
The network is composed by $J$ sensors and a fusion center.
We assume that each sensor $j \in \llbracket 1, J \rrbracket$ collects $N$ binary measurements $x_{j,n} \in \{0,1\}$ that are stored in a vector $\mathbf{x}_j$ of size $N$.
Consider $K$ different clusters $\mathcal{C}_k$ where each cluster is associated to a centroid $\boldsymbol{\theta}_k$ of length $N$. 
The binary components $\theta_{k,n}$ of $\boldsymbol{\theta}_k$ are independent and identically distributed (i.i.d.) with $P(\theta_{k,n} = 1) = p_c$.
We assume that each measurement vector  $\mathbf{x}_j$ belongs to one of the $K$ clusters.
The cluster assignment variables $e_{j,k}$ are defined as $e_{j,k}=1$ if $\mathbf{x}_j \in \mathcal{C}_k$, $e_{j,k}=0$ otherwise.
Let $\Theta = \{ \boldsymbol{\theta}_1,\cdots, \boldsymbol{\theta}_K\}$ and $E = \{ e_{1,1}, \cdots, e_{J,K}\}$ be the sets of centroids and of cluster assignment variables, respectively.
Within cluster $\mathcal{C}_k$, each vector $\mathbf{x}_j \in \mathcal{C}_k$ is generated as
\begin{equation}\label{eq:model_measurement}
 \mathbf{x}_j = \boldsymbol{\theta}_k \oplus \mathbf{b}_j,
\end{equation}
where $\oplus$ represents the XOR componentwise operation, and $\mathbf{b}_j$ is a vector of size $N$ with binary i.i.d. components  such that $P(b_{j,n} = 1) = p$.
We assume that the parameter $p$ and $p_c$ are unknown.
This model is equivalent to the model presented in~\cite{li05ACM} for K-means clustering with binary data.
It is symmetric, memoryless, and additive, which may not capture all the noise effect in practical situations, especially when the binary data comes from quantized real-valued data. 
Here, we consider this model as a first step to introduce the analysis and more accurate models will be considered in future works. 

The objective of the K-means algorithm is to recover the unknown cluster assignments $E$ and centroids $\Theta$.
Some instances of the K-means algorithm such as K-means++ have been proposed to deal with an unknown number of clusters $K$~\cite{jain10PRL}.
Here, as a first step, $K$ is assumed to be known in order to focus on the compression aspects of the problem.
In our context, each sensor has to transmit its data to the fusion center that should perform K-means on the received data. 
We now describe the source coding technique that is used in our system in order to reduce the amount of data transmitted to the fusion center.

\subsection{Source Coding with Sparse Binary Matrices}
In~\cite{xiong04SP}, it is shown that sparse binary matrices are very efficient to perform distributed source coding in a network of sensors, and in~\cite{toto11ComL,dupraz14Com} it is shown that they allow parameter estimation over the compressed data.
Denote by $H$ a binary matrix of size $ N \times M$ ($M < N$).
Denote by $d_{v} \ll M$ the number of non-zero components in any row of $H$, and denote by $d_{c} \ll N$ the number of non-zero components in any column of $H$.
In our system, each sensor $j$ transmits to the fusion center a binary vector $\mathbf{u}_j$ of length $M$, obtained as
\begin{equation}\label{eq:ujeq}
 \mathbf{u}_j = H^T \mathbf{x}_j ,
\end{equation}
where the operation is performed modulo $2$ and $T$ is the transpose operator applied to $H$. 
All the sensors use the same matrix $H$ with coding rate $r = \frac{M}{N}$.

The set of all the possible vectors $\mathbf{x}_j$ is called the original domain and is denoted as $\mathcal{X}^N = \{0,1\}^N$. 
The set of all the possible vectors $\mathbf{u}_j $ is called the compressed domain and is denoted by $\mathcal{U}^M \subseteq  \{0,1\}^M$.
The compressed domain $\mathcal{U}^M$ depends on the considered code $H$.
Note that in distributed source coding~\cite{xiong04SP}, the matrix $H$ is constructed as the sparse parity check matrix of a Low Density Parity Check (LDPC) code, which permits an efficient decoding of the vectors $\mathbf{x}_j$  at the fusion center.
Here, we do not want to reconstruct the original vectors $\mathbf{x}_j$, but the theoretical analysis carried in the paper will justify that the matrix $H$ still needs to be sparse in order to improve the performance of the K-means algorithm in the compressed domain.
The sparsity of $H$ also makes the encoding operation~\eqref{eq:ujeq} less complex, that is to say linear with the measurement vector length $N$. 



As in~\cite{xiong04SP}, we assume that the vectors $\mathbf{u}_j$ are transmitted reliably to the fusion center. 
We consider this assumption in order to focus on the source coding aspects of the problem, and we do not describe the channel codes that should be used in the system in order to satisfy this assumption.
Here, in order to avoid complex decoding operations as in~\cite{xiong04SP}, we propose to apply the K-means algorithm directly over the compressed vectors $\mathbf{u}_j$ received by the fusion center. 

\section{K-means Algorithm}\label{sec:method}
The K-means algorithm for clustering binary vectors $\mathbf{x}_j \in \mathcal{X}^N$ was initially proposed in~\cite{huang98DMKD}. 
In this section, we restate this algorithm in the compressed domain $\mathcal{U}^M$.
The Hamming distance between two vectors $\mathbf{a}$, $ \mathbf{b}$ $\in \mathcal{U}^M$ is defined as 
$
 \text{d}(\mathbf{a}, \mathbf{b}) = \sum_{m=1}^{M} a_m \oplus b_m .
$
Denote $\boldsymbol{\psi}_k = H^T \boldsymbol{\theta}_k$ and $\Psi = \{ \boldsymbol{\psi}_1,\dots, \boldsymbol{\psi}_K \}$ the compressed versions of the centroids $\boldsymbol{\theta}_k$.
Applying the K-means algorithm in the compressed domain corresponds to minimizing the objective function
$
 \mathcal{F}(\Psi, E) = \sum_{j=1}^J \sum_{k=1}^K e_{j,k} \text{d}(\mathbf{u}_j, \boldsymbol{\psi}_k) .
$
with respect to the compressed centroids $\boldsymbol{\psi}_k$ and to the assignment variables $e_{j,k}$.

We initialize the K-means algorithm with $K$ compressed centroids $\boldsymbol{\psi}_k^{(0)}$ that may be either selected at random among the set of input vectors $\mathbf{u}_j$, or obtained from the K-means++ procedure~\cite{arthur07SIAM}.
Denote by $L$ the number of iterations of the K-means algorithm. 
In the following, superscript $\ell$ always refers to a quantity obtained at the $\ell$-th iteration of the algorithm.
At iteration $\ell \in \llbracket 1,L \rrbracket$, K-means proceeds in two steps.
First, from the centroids $\boldsymbol{\psi}_k^{(\ell-1)}$ obtained at iteration $\ell-1$, it assigns each vector $\mathbf{u}_j$ to a cluster as
\begin{equation}\label{eq:cluster_transform}
 \forall j, k, e_{j,k}^{(\ell)} = \left\{ \begin{array}{ll}
                               1 \text{ if } d(\mathbf{u}_j, \boldsymbol{\psi}_k^{(\ell-1)}) = \underset{ k' \in \llbracket 1,K\rrbracket}{\min} d(\mathbf{u}_j, \boldsymbol{\psi}_{k'}^{(\ell-1)}),  \\
                               0 \text{ otherwise.}
                              \end{array}
 \right.
\end{equation}
Second, the algorithm updates the centroids as follows:
\begin{equation}\label{eq:centroid_transform}
 \forall j, n, ~~ \psi_{k,n}^{(\ell)} = \left\{ \begin{array}{ll}
					   1 & \text{ if } \overset{J}{\underset{j=1}{\sum}} e_{j,k}^{(\ell)} u_{j,n} \geq \frac{1}{2} J_k^{(\ell)}, \\
					   0 & \text{otherwise}.
					  \end{array} \right.
\end{equation}
where $J_k^{(\ell)}$ is the number of vectors assigned to cluster $k$ at iteration $\ell$.
Step~\eqref{eq:cluster_transform} assigns each vector $\mathbf{u}_j$ to the cluster with the closest compressed centroid $\boldsymbol{\psi}_k^{(\ell)}$.
The centroid computation step~\eqref{eq:centroid_transform} is a majority voting operation which can be shown to minimize the average distances between the centroid $\boldsymbol{\psi}_k^{(\ell)}$ and all the vectors $\mathbf{u}_j$ assigned to cluster $k$ at iteration~$\ell$.

Following the same reasonning as for K-means in the original domain~\cite{huang98DMKD}, it is easy to show that when applying K-means in the compressed domain, the sequence of objective functions $ \mathcal{F}(\Psi^{(\ell)}, E^{(\ell)})$ 
is decreasing with $\ell$ and converges to a local minimum.
However, this property does not guarantee that the cluster assignment variables $e_{j,k}^{(\ell)}$ obtained from the algorithm in the compressed domain will correspond to the correct cluster assignments in the original domain.
In order to show that the K-means algorithm applied in the compressed domain can recover the correct clusters of the original domain, we now propose a theoretical analysis of the two steps of the algorithm.

\section{K-means Performance Evaluation}\label{sec:perf}
In order to assess the performance of the K-means algorithm in the compressed domain, we evaluate each step of the algorithm individually.
We provide an approximated expression of the error probability of the cluster assignment step in the compressed domain, assuming that the compressed centroids $\boldsymbol{\psi}_k$ are perfectly known.
In the same way, we provide an approximated expression of the error probability of the centroid estimation step in the compressed domain, assuming that the cluster assignment variables $e_{j,k}$ are perfectly known.
Although evaluated in the most favorable cases, these error probabilities will enable us determine whether it is reasonable to apply K-means in the compressed domain in order to recover the clusters of the original domain. 
%
The expressions of the error probabilities we derive rely on three functions $f_M$, $\mathcal{F}_M$, and $g$ defined as
\begin{align}\label{eq:proba_ak}
 f_M(m,p) & = \binom{M}{m} p^{m} (1-p)^{M-m}, \\
 \mathcal{F}_M(m,p) & = \sum_{u=m+1}^M f_M(u,p), \\
 g(d,p) & =\frac{1}{2} - \frac{1}{2}(1-2p)^{d} .
\end{align}
The function $f_M$ is the Binomial probability distribution and the function $\mathcal{F}_M$ is the Binomial complementary cumulative probability distribution. The value $g(d,p)$ gives the probability that the sum of $d$ binary random variables $\sum_{i=1}^d X_i$ equals $1$, where $P(X_i=1)=p$, see~\cite[Section 3.8]{gallager62PhD}. 

 \subsection{Error Probability of the Cluster Assignment Step}\label{sec:perfcluster}
The following proposition evaluates the error probability of the cluster assignment step~\eqref{eq:cluster_transform} applied to the compressed centroids $\boldsymbol{\psi}_k$.

\begin{prop}\label{prop:cluster}
Let $\hat{e}_{j,k}$ be the cluster assignments obtained when applying the cluster assignment step~\eqref{eq:cluster_transform} to the true compressed centroids $\boldsymbol{\psi}_k$.
The error probability $P_{a,k}= P(\hat{e}_{j,k}=0|\mathbf{x}_j \in \mathcal{C}_k)$ for cluster $k$ can be approximated as
 \begin{equation}\label{eq:pe_clustera}
P_{a,k} \approx 1 - \sum_{m_1=0}^{M} \sum_{m_2=m_1}^{M}  f_M\left(m_1,q_1\right) B_{M,K}(m_2,q_2) 
 \end{equation}
 where \small
 \begin{equation}
   B_{M,K}(m_2,q_2) = \sum_{k=1}^K \binom{K-1}{k} f_M(m_1,q_2)^k \mathcal{F}_M(m_1,q_2)^{K-1-k} 
 \end{equation}

 and $ q_1  = g(d_c,p)$, $q_2  = g\left(d_c,\frac{1}{2}(1-(1-2p_c)^2(1-2p))\right)$ .
\end{prop}
\begin{proof}
 See appendix~\ref{app:Prop1}.
\end{proof}
It can be seen from~\eqref{eq:pe_clustera} that the approximated error probability $P_{a,}$ depends on the number of clusters $K$ but does not depend on the considered cluster $k$. 
The expressions~\eqref{eq:pe_clustera} and~\eqref{eq:error_centroid}  are only approximations of the error probabilities of the two steps of the algorithm. Indeed, they are obtained by assuming that the components of the vector $H^T \mathbf{b}_j $ are independent, which is not true in general.
However, it is shown in~\cite{dupraz14Com,toto11ComL} that this assumption is reasonable for parameter estimation over sparse binary matrices.
In Section~\ref{sec:results}, we verify the accuracy of this approximation by comparing the values of~\eqref{eq:pe_clustera} and~\eqref{eq:error_centroid} to the error probabilities measured from Monte Carlo simulations. 
 \subsection{Error Probability of the Centroid Computation Step}
 The following proposition now evaluates the error probability of the centroid computation step~\eqref{eq:centroid_transform} in the compressed domain.
 
 \begin{prop}\label{prop:centroids}
 Let $\hat{\Psi}_k$ be the estimated compressed centroids obtained after applying the centroid estimation step~\eqref{eq:centroid_transform} to the true cluster assignment variables $ e_{j,k}$.
 The error probability $P_{c,k} = P(\hat{\psi}_{k,m} \neq \psi_{k,m})$ for cluster $k$ can be approximated as
  \begin{equation}\label{eq:error_centroid}
 P_{c,k} \approx \sum_{j = \lceil\frac{J_k}{2}\rceil}^{J_k} f_{J_k}(j,q_1) 
\end{equation}
where $J_k$ is the number of vectors in cluster $k$, and $q_1 = g(d_c,p)$. 
 \end{prop}
\begin{proof}
 See appendix~\ref{app:Prop2}.
\end{proof}
The approximated error probability $P_{c,k}$~\eqref{eq:error_centroid} only depends on the considered cluster $k$ through the number $J_k$ of vectors in cluster $\mathcal{C}_k$.
The expression~\eqref{eq:pe_clustera} is only an approximation of the error probability of the centroid assignment step for the same reasons as for the cluster assignment step.
We will also verify the accuracy of this approximation in Section~\ref{sec:results}.

\section{Simulation Results}\label{sec:results}
\begin{figure*}[t]
\begin{center}
  \subfigure[~]{ \includegraphics[width=.45\linewidth]{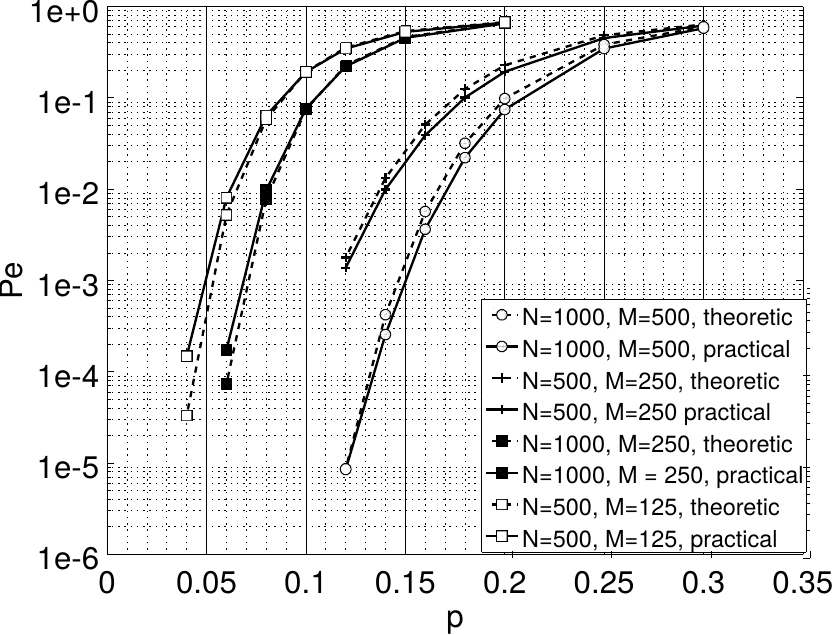}}
  \subfigure[~]{ \includegraphics[width=.45\linewidth]{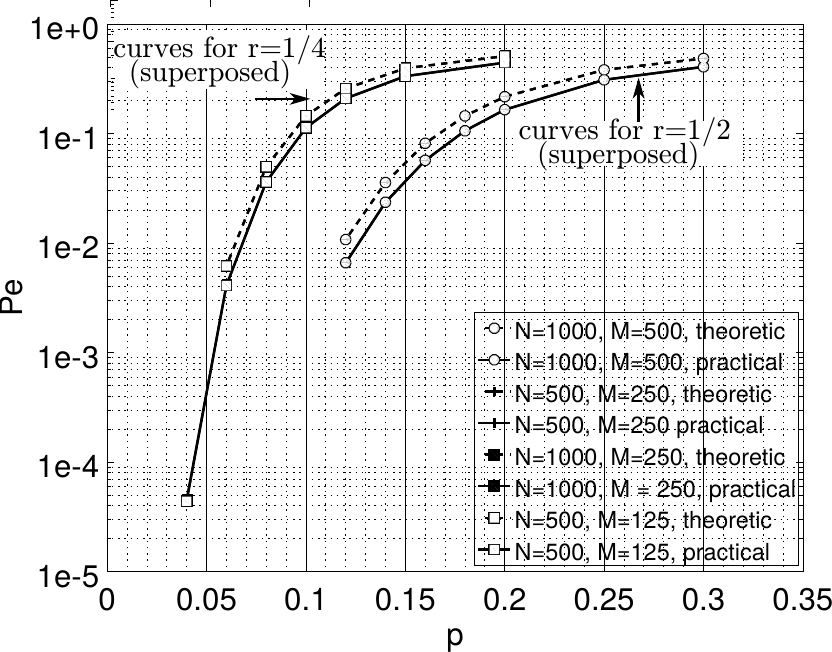}}
\end{center}
\caption{Comparison of approximated theoretic error probabilities with error probabilities measured from Monte Carlo simulations for the two steps of the algorithm. For both steps, dashed lines represent theoretic error probabilities while continuous lines give error probabilities measured from Monter Carlo simulations. (a) Cluster assignment step (b) Centroid estimation step. }
\label{fig:allfigs}
\end{figure*}

\begin{figure}[t]
 \begin{center}
  \includegraphics[width=.4\linewidth]{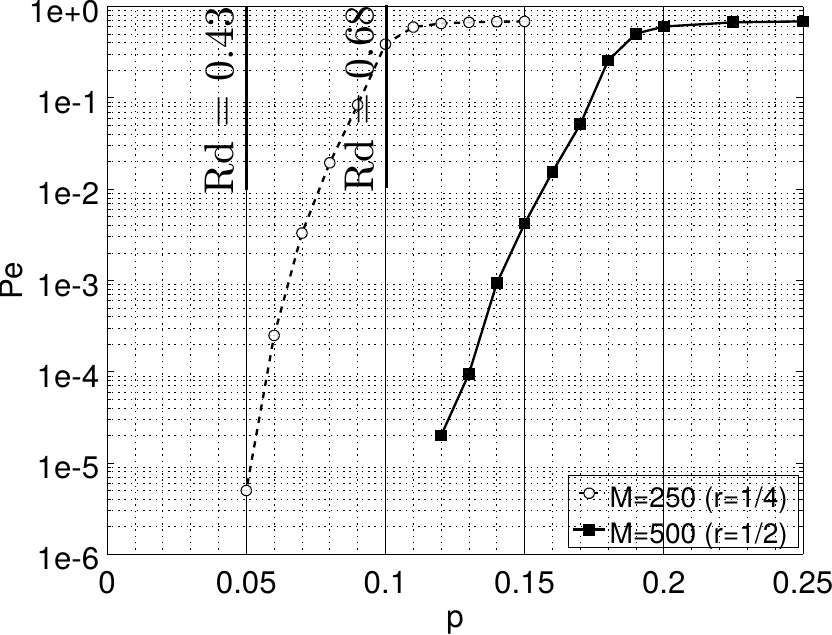}
  \caption{  Error probability of the K-means algorithm with $L=10$ iterations. The values of $R_d$ indicate the average rates needed to reconstruct all the measurement vectors for the considered values of $p$. \vspace{-0.5cm} }
  \label{fig:perf}
 \end{center}
\end{figure}

In this section, we evaluate through simulations the performance of the K-means algorithm in the compressed domain.
We first consider each step of the algorithm individually, and we verify the accuracy of the approximated error probabilities obtained in Section~\ref{sec:perf}.
We then assess the performance of the full algorithm and we evaluate the rate needed to perform K-means over compressed data.

Throughout the section, we set $J=200$, $K=4$, $p_c=0.1$. We set $d_v=2$ for all the considered binary sparse matrices, since it can be shown from~\eqref{eq:pe_clustera} and~\eqref{eq:error_centroid} that the error probabilities $P_a$ and $P_{c,k}$ are increasing with $d_v$ ($d_v$ is necessarily greater than $2$). 
The sparse matrices are constructed from the Progressive Edge Growth algorithm~\cite{hu05IT}, which reduces the correlation between the components of $H^T \mathbf{b}_j$. 
%

\vspace{-0.3cm}

\subsection{Accuracy of the error probability approximations}
Here, we consider two codes of rate $r=1/2$ and $d_c=4$ with parameters $(N=1000, M=500)$ and $(N=500,M=250)$. We also consider two codes of rate $r=1/4$ with $d_c=8$ and parameters $(N=1000, M=250)$ and $(N=500,M=125)$.
We compare the approximated expressions $P_a$~\eqref{eq:pe_clustera} and $P_{c,k}$~\eqref{eq:error_centroid} with the effective error probabilities measured from Monte Carlo simulations for each step of the algorithm for the four constructed codes over $Nt=10000$ simulations.
Figure~\ref{fig:allfigs}(a) represents the obtained error probabilities for the cluster assignment step, while Figure~\ref{fig:allfigs}(b) represents the centroid estimation step.
As expected from~\eqref{eq:pe_clustera} and~\eqref{eq:error_centroid}, the performance of the cluster assignment step varies with the vector length $N$, but the performance of the centroid computation step does not vary with $N$. 

We also see that for each of the two steps of the algorithm, the theoretic error probabilities $P_{a,k}$ and $P_{c,k}$  are very close to the measured error probabilities, whatever the considered code. This shows the accuracy of the proposed approximations, even for smaller values $N=500$ for which the correlation between components of $H^T\mathbf{b}_j$ increases.
Figure~\ref{fig:allfigs}(a) and (b) also illustrate that the cluster assignment step and the centroid computation step in the compressed domain can indeed recover the correct clusters of the original domain, since it is possible to reach error probabilities from $10^{-3}$ to $10^{-7}$.

\vspace{-0.3cm}

\subsection{K-means algorithm and rate evaluation}
In this section, we evaluate the performance of the K-means algorithm in the compressed domain. 
Here, we set $N=1000$. We consider a first code of rate $r=1/4$ with $M=250$ and $d_c=8$, and a second code of rate $r=1/2$, with $M=500$ and $d_c=4$.
We run  over $Nt=10000$ simulations the K-means algorithm in the compressed domain initialized with $L=10$ iterations. For each simulation, the K-means algorithm is repeated $100$ times over the same data in order to avoid initialization issues. 
Figure~\ref{fig:perf} represents the error probability of the cluster assignments decided by the algorithm in the compressed domain with respect to the correct clusters in the original domain.
The error probability is decreasing with $r$ and is increasing with $p$, which is expected since the value of $p$ represents the noise level in the measurement vectors compared to the centroids.

The above results show that the rate $r$ has to be chosen carefully with respect to the value of $p$ in order to guarantee the efficiency of the clustering in the compressed domain.
It can be shown from the value of $R_d$ that the rate needed to reconstruct all the measurement vectors also increases with $p$. 
This is why we now compare the rate needed to perform K-means over compressed data to the rate needed to reconstruct all the sensors measurements.
For $p_c=0.1$ and $p=0.1$, we get $R_d=0.68$ bits/symbol, and, for $p_c = 0.1$ and $p=0.05$, we obtain $R_d = 0.43$ bits/symbol.
The results of Figure~\ref{fig:perf} show that for $p_c=0.1$ and $p=0.1$, the code of rate $r=1/2 < 0.68$ enables to perform K-means with an error probability lower than $10^{-6}$.
For $p_c=0.1$ and $p=0.05$, the code of rate $r=1/4 < 0.43$ also enables to perform K-means with a low error probability $P_e=10^{-5}$.
This shows that the rate needed to perform K-means is lower than the rate needed to reconstruct all the sensors measurements, which justifies the approach presented in the paper.

\section{Conclusion}\label{sec:conclusion}
In this paper, we considered a network of sensors that transmit their compressed binary measurements to a fusion center. 
We proposed to apply the K-means algorithm directly over the compressed data, without reconstructing the sensor measurements.
From a theoretical analysis and from Monte Carlo simulations, we showed the efficiency of applying K-means in the compressed domain.
We also showed that the rate needed to perform K-means on the compressed vectors is lower than the rate needed to reconstruct all the measurements.
%
Future works will also be dedicated to the generalization of the analysis to more complex, \emph{e.g.} asymmetric models or models with memory.

\appendix
\Section{Appendix}
\subsection*{Proof of Proposition 1}\label{app:Prop1}
Without loss of generality, we first evaluate the error probabilities $P_{a,1}= P(\hat{e}_{j,1}=0|\mathbf{x}_j \in \mathcal{C}_1)$.
Assume that $\mathbf{x}_j \in \mathcal{C}_1$ and let
\begin{align}\notag
 \mathbf{a}_1 & = \mathbf{u}_j \oplus \boldsymbol{\psi}_1 =  H^T \mathbf{b}_j, \\
 2\leq k\leq K, ~ \mathbf{a}_{k} & = \mathbf{u}_j \oplus \boldsymbol{\psi}_{k} =  H^T ( \boldsymbol{\theta}_{1} \oplus \boldsymbol{\theta}_{k} \oplus \mathbf{b}_j ) .
\end{align}
Define for all $k\in \{1,\cdots, K\}$, $A_i = \sum_{m=1}^M a_{k,m}$.
According to the cluster assignment step~\eqref{eq:cluster_transform}, the error probability $P_{a,1}$ can be expressed as 
\begin{equation} \label{eq:Pa1}
 P_{a,1} =  1 -  P\left(  A_1 \leq \min_{i=2,\cdots, K} A_i \right) \approx 1 - \sum_{u=0}^M P(A_1 = u) \sum_{v=u}^M P\left( \min_{i=2,\cdots, K} A_i = v\right) .
\end{equation}
In the above expression, the probability of the minimum of the $A_i$ can be developped as
\begin{align}\label{eq:Pmin}
 P \left(  \min_{i=2,\cdots, K} A_i = v\right)  \approx  \sum_{k=1}^{K-1} \binom{K-1}{k} P(A_i = v)^k P(A_i>v)^{K-1-k}
\end{align}
given that the $A_i$ with $i>1$  are all identically distributed.
 In order to get~\eqref{eq:Pa1} and~\eqref{eq:Pmin} we implicitly assume that the random variables $A_i$ are mutually independent for all $i=1,\cdots,K$, and as a result~\eqref{eq:Pa1} is only an approximation of $P_{a,1}$. 
 To finish, the terms $P(A_1=u)$ and $P(A_i = v)$ ($i\neq 1)$ can be calculated as follows. 
 First, since $a_{1,m}$ is the XOR sum of $d_c$ binary random variables $b_{j,n}$, its probability is given by $P(a_{1,m}=1) = q_1$. 
 Assuming that the $a_{1,m}$ are independent, it follows that $P(A_1=u) \approx f_M(u,q_1)$. 
 In the same way, for $i>1$, $P(a_{i,m}=1) = q_2$ since $a_{i,m}$ is the XOR sum of $d_c$ random variables $\theta_{1,n} \oplus \theta_{i,n} \oplus b_{j,n}$ and $P(\theta_{i,n} = 1) = p_c$.
 This gives $P(A_i=v) \approx f_M(v,q_2)$.
 At the end, the terms $P(A_i>v)$ are given by the Binomial complementary cumulative probability distribution $\mathcal{F}_M(v,q_2)$. 


 
\subsection*{Proof of Proposition 2}\label{app:Prop2}
 From the model defined in Section~\ref{sec:system}, a codeword $\mathbf{u}_j$ ($j \in \mathcal{C}_k$), can be expressed as
\begin{equation}
 \mathbf{u}_j = H^T( \boldsymbol{\theta}_k \oplus \mathbf{b}_j) = \boldsymbol{\psi}_k \oplus \mathbf{a}_j,
\end{equation}
where $\mathbf{a}_j = H^T\mathbf{b}_j$ is such that $P(a_{j,m}=1) = p_d $.
Let $A_j = \sum_{j=1}^{J_k} a_{j,m}$.
The error probability of the centroid computation step can be evaluated as
\begin{equation}\label{eq:Pecentroid}
 P_{c,k} = P\left( A_j \geq \frac{J_k}{2} \right) \approx \sum_{j = \lceil\frac{J_k}{2}\rceil}^{J_k} f_{J_k}(j,q_1) .
\end{equation}
The approximation comes from the fact that~\eqref{eq:error_centroid} assumes that the $a_{j,m}$ are independent.

\Section{References}
\bibliographystyle{ieeetr}
\bibliography{biblio}

\end{document}